\documentclass[usenatbib,useAMS,times]{mn2e}
\usepackage{graphicx}

\usepackage{aas_macros}
\title[]{Observations of Ultracool dwarfs with {\sc ultracam} on the {\sc
vlt}: a search for weather$^1$}

\author[]{S.\,P.\ Littlefair$^{1}$, V.\,S.\, Dhillon$^{1}$, 
T.\,R.\, Marsh$^{2}$, T.\, Shahbaz$^{3}$, E.\,L.\, Mart\'{i}n$^{3,4}$\\
$^1$Dept of Physics and Astronomy, University of Sheffield, Sheffield, S3 7RH, UK \\
$^2$Dept of Physics, University of Warwick, Coventry, UK\\
$^3$Instituto de Astrof\'{i}sica de Canarias, Tenerife, Spain\\
$^4$University of Central Florida, Dept. of Physics, PO Box 162385, Orlando, FL 32816-2385, USA\\}

\date{\center{\Large Submitted for publication in the Monthly Notices of the
Royal Astronomical Society \\ 
\vspace{.5cm} \today}} 

\begin{document}
\maketitle

\begin{abstract} 
We present multi-colour photometry of four field ultracool dwarfs with
the triple-beam photometer {\sc ultracam}. Data were obtained
simultaneously in the Sloan-$g'$ band and a specially designed
narrow-band NaI filter. The previously reported 1.8-hr period of
Kelu-1 is here recovered in the $g'$-band, but the lack of any
significant variability in the NaI light of this object precludes any
conclusion as to the cause of the variability. 2MASS 2057-0252 and
DENIS 1441-0945 show no convincing evidence for variability. 2MASS
1300+1912, on the other hand, shows good evidence for gradual trends in
both bands at the 5\% level. These trends are anti-correlated at a
high level of significance, a result which is incompatible with models
of starspot-induced variability. It would seem likely that dust cloud
``weather'' is responsible for the short-term variability in this
object.

\end{abstract} 


\section{Introduction}
\label{sec:introduction}
\alph{footnote} \protect\footnotetext[1]{Based on observations made at
the European Southern Observatory, Paranal, Chile (ESO program
075.C-0505)} The photometric variability of brown dwarfs and very
low-mass stars (collectively known as ultracool dwarfs or UCDs) has
received substantial attention over the last five years. The recent
review by \citet{bailer-jones04} suggests that around 40\% of UCDs
which are surveyed show variability. The variability appears to be
transient in nature - for example, observations of the L3 dwarf 2MASS
1146+2230 by \citet{gelino02} and \citet{clarke02a} failed to find the
variability reported by \citet{bailer-jones01}.  Also, extensive
observations of the M9.5 dwarf BRI 0021-0214 by \citet{martin01}
found both periodic and transient $I$-band variability that did not
correspond to the expected rotation period. It is therefore possible
that {\em all} UCDs show variability at some level.

The variability falls into two broad categories: periodic modulations
with periods up to a few days and magnitudes of up to 100 mmag, and
non-periodic variations with timescales on the order of hours to days and
amplitudes of $\sim10$ to $\sim100$ mmag in the $I$-band, where most
of the observations to date have been undertaken. The existence of
non-periodic variability is interesting, as the monitoring surveys
undertaken would have been sensitive to the rotation periods of UCDs.
The existence of non-periodic variability led \citet{bailer-jones01}
to propose that the variability is due to surface features which 
evolve on timescales shorter than the rotational period.

There are two prime candidates for the surface features responsible
for variability. Cool, magnetic star-spots are an attractive
explanation for the very young UCDs observed in clusters. Star-spots
are known to cause variability in the more massive T Tauri stars
\cite[see][for example]{herbst94}, and furthermore, young brown dwarfs
show similar X-ray properties to T-Tauri stars, implying that a
similar magnetic activity mechanism is at work in young UCDs and T
Tauri stars \cite[e.g.][]{ozawa05,bouy04}. The older, field, UCDs may
not show magnetic spots, as the cooler photospheres will be neutral
and result in a weak coupling between the gas and magnetic field
\citep{mohanty02,fleming02}. For the field UCDs, dust clouds are a more
plausible candidate: dust should form under the cool and dense
conditions found in their photospheres, and dusty atmospheric models
are more successful in reproducing the observed colors of UCDs than
dust-free models \citep{leggett98,martin00}. The atmospheres of UCDs
are also likely to be highly dynamic, as a result of turbulence
excited by convection \citep{allard01}. It is quite
likely that rapidly evolving dust cloud ``weather'' is responsible for
the non-periodic variability seen in UCDs.

The different spectral signatures of variability caused by spots and
dust clouds allows one, in principle, to distinguish between these two
candidates based upon time-resolved spectrophotometry. Unfortunately,
the results so far have been inconclusive, as no significant
variability was detected in dust-sensitive wavelength regions
\citep{clarke03,bailer-jones02}.  The reasons for this lack of success
are not clear. It may be simply that variability in UCDs is transient,
and the limited number of observations to date have merely been
unlucky. Alternatively, it may be that the systematic errors in
spectrophotometry, arising from sky subtraction, telluric correction
and slit-loss correction are too large to attain the required
accuracies of a percent or better implied by the magnitude of
variability in the $I$-band.

\begin{figure}
\begin{center}
\includegraphics[scale=0.3,angle=90,trim=50 0 0 80]{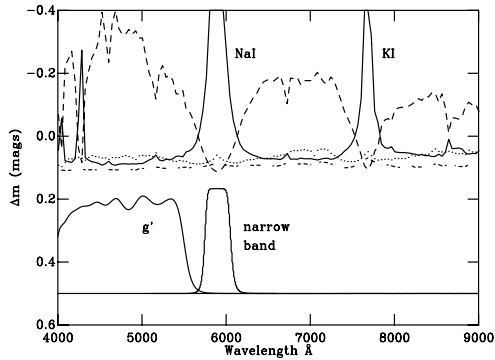}
\caption{Calculations of the {\em change} in the spectrum of a UCD
with $T_{eff} = 1900$K, due to the appearance of a cloud or spot which
covers 10\% of the stellar surface. The calculations were made using
the synthetic spectra of \protect\cite{allard01}. The four lines
represent a the change in the spectrum of the UCD, caused by the
formation of: a clear hole in a dusty atmosphere (dashed line), a
dusty cloud on a clear atmosphere (solid line), a 200K cooler spot on
a dusty atmosphere (intermittent dashed line) and a 200K cooler spot
on a clear atmosphere (dotted line). Also shown are the bandpasses of
the Sloan-$g'$ filter, and the specially designed NaI filter
($\lambda_{cen} = 5900$\AA$, \delta\lambda = 220$\AA) used for the
observations presented in this paper.}
\label{fig:model}
\end{center}
\end{figure}
A better approach may be to use time-resolved photometry, taken
simultaneously in a narrow band centred on dust-centred features and a
continuum filter. Photometry does not suffer from slit-loss errors, as
other stars in the field can ve used to correct for the light that
falls outside the aperture. Also, the photons are collected in fewer
pixels, allowing greater signal to noise ratios to be achieved.
Narrow band photometry was used by \cite{tinney99}, who obtained
quasi-simultaneous narrow band photometry of two UCDs, DENIS-P
J1228-1547 and LP 944-20. Variability was found in both bands for LP
944-20, showing the promise of photometry for future
studies. Unfortunately, the complex nature of the spectral behaviour
in the narrow bands chosen by \cite{tinney99}, precluded any firm
conclusion as to the origin of the variability.  We outline here a
modification of the method proposed by \cite{tinney99}. Multi-band
photometry has also been used succesfully to show that, in some cases
at least, the variability of M-dwarfs is due to magnetic spots
\citep{rockenfeller06a}.  Figure~\ref{fig:model} shows the details of
the method (which is described in more detail by
\citealt{bailer-jones01}). The opacity within a spot feature shows a
broadly similar colour dependence to the opacity of the immaculate
photosphere, but the cool spot lowers the average temperature of the
visible portion of the star. Therefore, the evolution or appearance of
a spot feature produces a dimming of the spectrum at all
wavelengths. If we choose to observe simultaneously in two filters,
one centred on the continuum, the other centred on the neutral alkali
lines of NaI, a cool spot will cause magnitude changes which are {\em
correlated} between the filters. In the case of the formation of a
clearing within a dusty cloud deck, the clearing {\em removes} opacity
in the continuum, but significantly {\em increases} opacity in
spectral regions dominated by molecular lines (the strong resonance
lines of NaI seen at 0.59 $\mu$m are good examples of such a
region). A dust cloud forming on a clear photosphere has the opposite
effect, increasing the continuum opacity, but decreasing the opacity
at NaI wavelengths. Thus, in contrast to the case of a starspot, a
change in the cloud coverage of a UCD causes {\em anti-correlated}
magnitude changes between the two filters. By centering our narrow
band filter on the neutral alkali lines, we can thus distinguish
between spots and dust clouds.

In this paper we present multi-colour photometry of four field
ultracool dwarfs with the triple-beam photometer {\sc ultracam}
\citep{dhillon01}. Data were taken simultaneously in the Sloan-$g'$
band and the specially designed narrow-band NaI filter shown in
figure~\ref{fig:model}. The observations are described in
section~\ref{sec:obs}, the results presented in
section~\ref{sec:results} and discussed in section~\ref{sec:disc},
whilst in section~\ref{sec:conc} we draw our conclusions.

\section{Observations}
\label{sec:obs}
The UCDs 2MASS 1300+1912, Kelu-1, DENIS 1441-0945 and 2MASS 2057-0252
were observed simultaneously in the Sloan-$g'$ and NaI bands using
{\sc ultracam} on the 8.2-m {\sc melipal} unit of the Very Large
Telescope (VLT) at Paranal, Chile. Thanks to the frame-transfer CCDs
employed in {\sc ultracam} we were able to obtain 10 second exposures
with no dead-time between frames. Sloan-$u'$ images were obtained
simultaneously with the other colours, but the UCDs are too faint to
produce useful lightcurves in this band. The observations are
summarised in Table~\ref{table:obs}.  Data reduction was carried out
using the {\sc ultracam} pipeline date reduction software. Extraction
of target and comparison star lightcurves was performed using an
optimal extraction method \citep{naylor98}. Due to the 2.6\arcmin
field of view of {\sc ultracam} on the VLT and the sparse nature of
the target fields, a limited number of comparison stars are available
for each object.

On inspection of the images, it is clear that the images suffer from
structured vignetting and scattered light components at the 10\%
level. Further investigation of these effects suggest that they arise
in the collimator optics. To prevent the scattered light in the
flat-field frames from introducing an error in the flat-field process,
large-scale trends were removed from the flat-field before use, by
dividing the flat-field by a median-filtered version of itself. This
means that vignetting is {\em not} corrected for by our flat field.
For the majority of our observations this is not a major problem, as
the position of the stars on the chip remained stable to within 1
pixel throughout observation.  Whilst the absolute flux level for
these stars will be in error because of the lack of vignetting
correction, the shape of the light-curves will not be affected. The
one exception to this is the observation of 2MASS 1300+1912.  At
various times in the observation of this object, the target was moved
around the chip. In principle this might introduce spurious features
into the light curve, as the relative vignetting between target and
comparison star changes.  In practice, the distance between comparison
star and target was small ($\sim$140 pixels), and we believe this
effect is negligible, a conclusion which is reinforced by the relative
flatness of the comparison star lightcurve in
figure~\ref{fig:lightcurves}, and the lack of sudden jumps in the
lightcurves at times when the star was moved.
\begin{table}
\caption[]{Journal of Observations.}
\begin{center}
\begin{tabular*}{3.5in}{@{\extracolsep{-0.2cm}}llcccc}
& & & & & \\

\multicolumn{1}{c}{UT Date} & \multicolumn{1}{c}{Target}&
\multicolumn{1}{c}{UT} & \multicolumn{1}{c}{UT} &
\multicolumn{1}{c}{Seeing} & Airmass \\ \multicolumn{1}{c}{Start} & &
\multicolumn{1}{c}{Start} & \multicolumn{1}{c}{End} &
\multicolumn{1}{c}{arcsecs} & \\ & & & & & \\ \hline \hline & & & & &
\\ 2005-05-08 & 2MASS 1300+1912 & 23:22 & 04:13 & 0.6--1.0 & 1.4--2.3
\\ 2005-05-13 & DENIS 1441-0945 & 03:11 & 07:02 & 0.5--0.8 & 1.0--1.4
\\ 2005-05-13 & 2MASS 2057-0252 & 07:05 & 10:14 & 0.5--0.8 & 1.1--1.5
\\ 2005-05-14 & Kelu-1 & 02:37 & 06:49 & 0.5--1.0 & 1.0--2.0 \\
2005-05-17 & 2MASS 2057-0252 & 07:52 & 09:50 & 0.8--1.1 & 1.0--1.1 \\
& & & & & \\ \hline & & & & & \\
\end{tabular*}
\end{center}
\label{table:obs}
\end{table}

\section{Results}
\label{sec:results}

\subsection{Rapid Variability}
We performed a search for rapid variability in all our objects, by
calculating the reduced $\chi$-squared with respect to the polynomial
fits shown in figure~\ref{fig:lightcurves}. In each case, the reduced
$\chi$-squared with rspect to the fits was less than or equal to a
value of 1.5. We conclude that no objects show evidence for ``rapid''
variability, i.e variability on timescales of a few minutes. In the
rest of the paper, the term variability refers to medium term trends
within the data, i.e variability on the timescale of hours.

\subsection{2MASS 2057-0252}
\label{subsec:2mass2057}

\begin{figure*}
\begin{center}
\includegraphics[scale=0.35,trim=-10 -10 -10 -10]{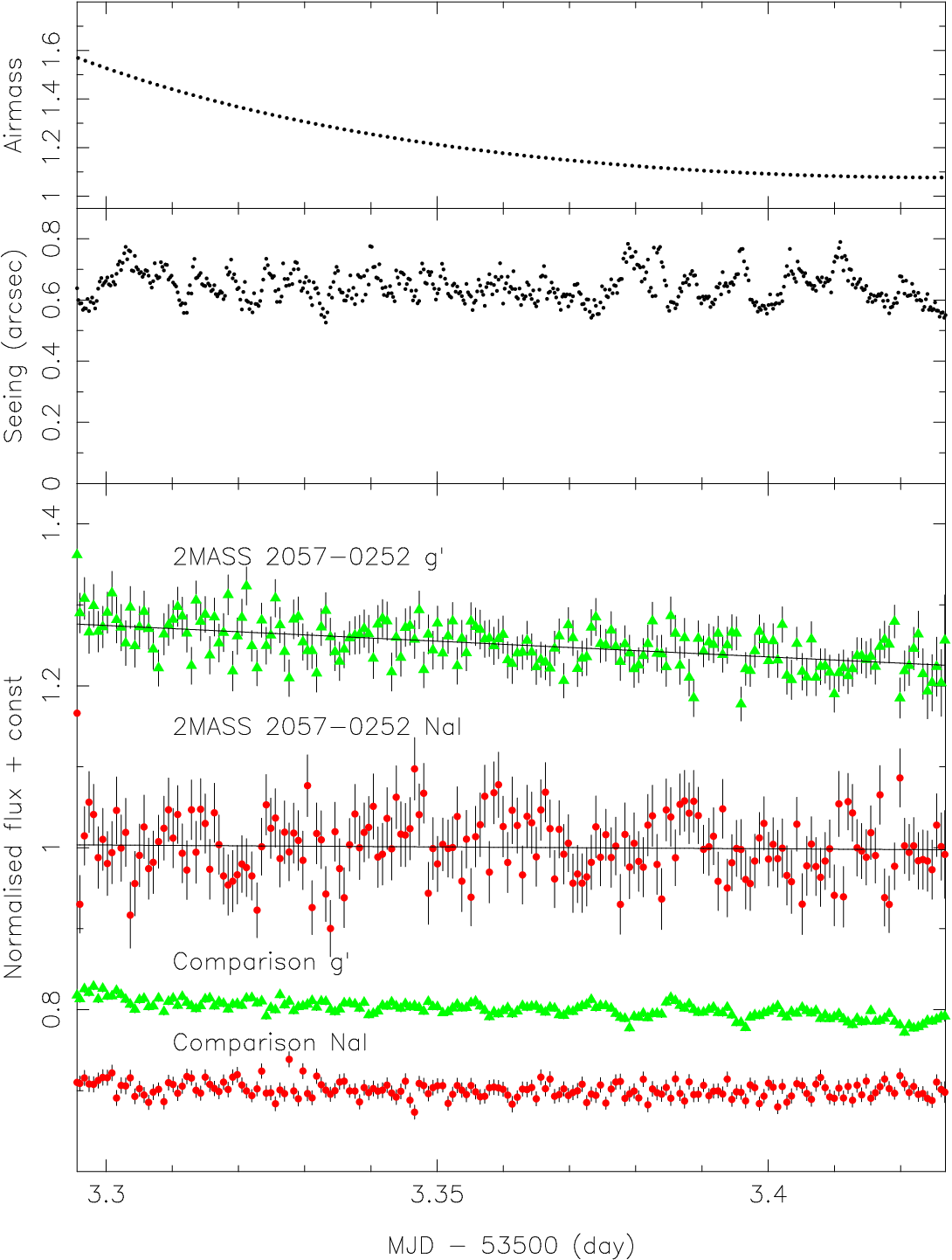}
\includegraphics[scale=0.35,trim=-10 -10 -10 -10]{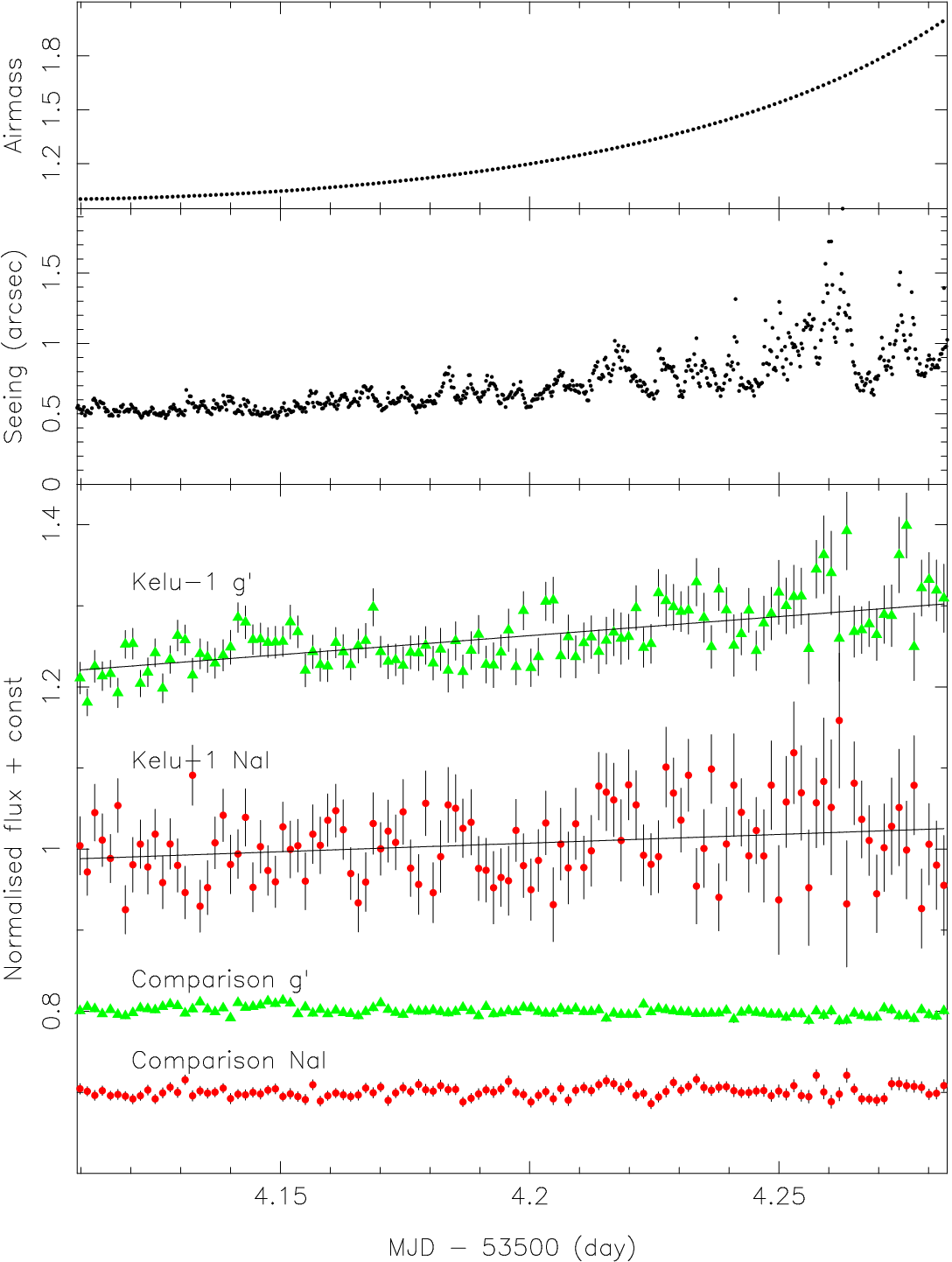}
\includegraphics[scale=0.35,trim=-10 -10 -10 -10]{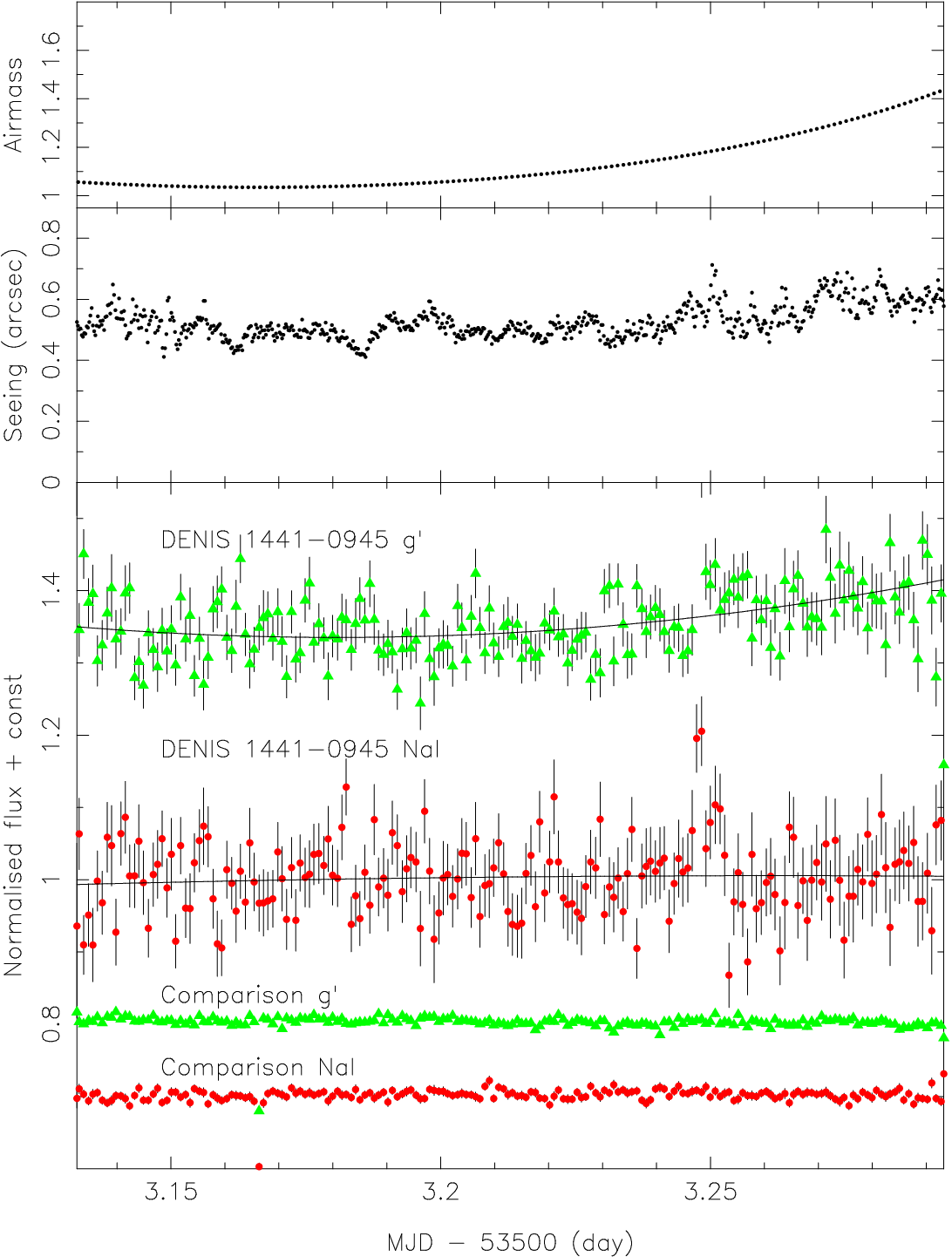}
\includegraphics[scale=0.35,trim=-10 -10 -10 -10]{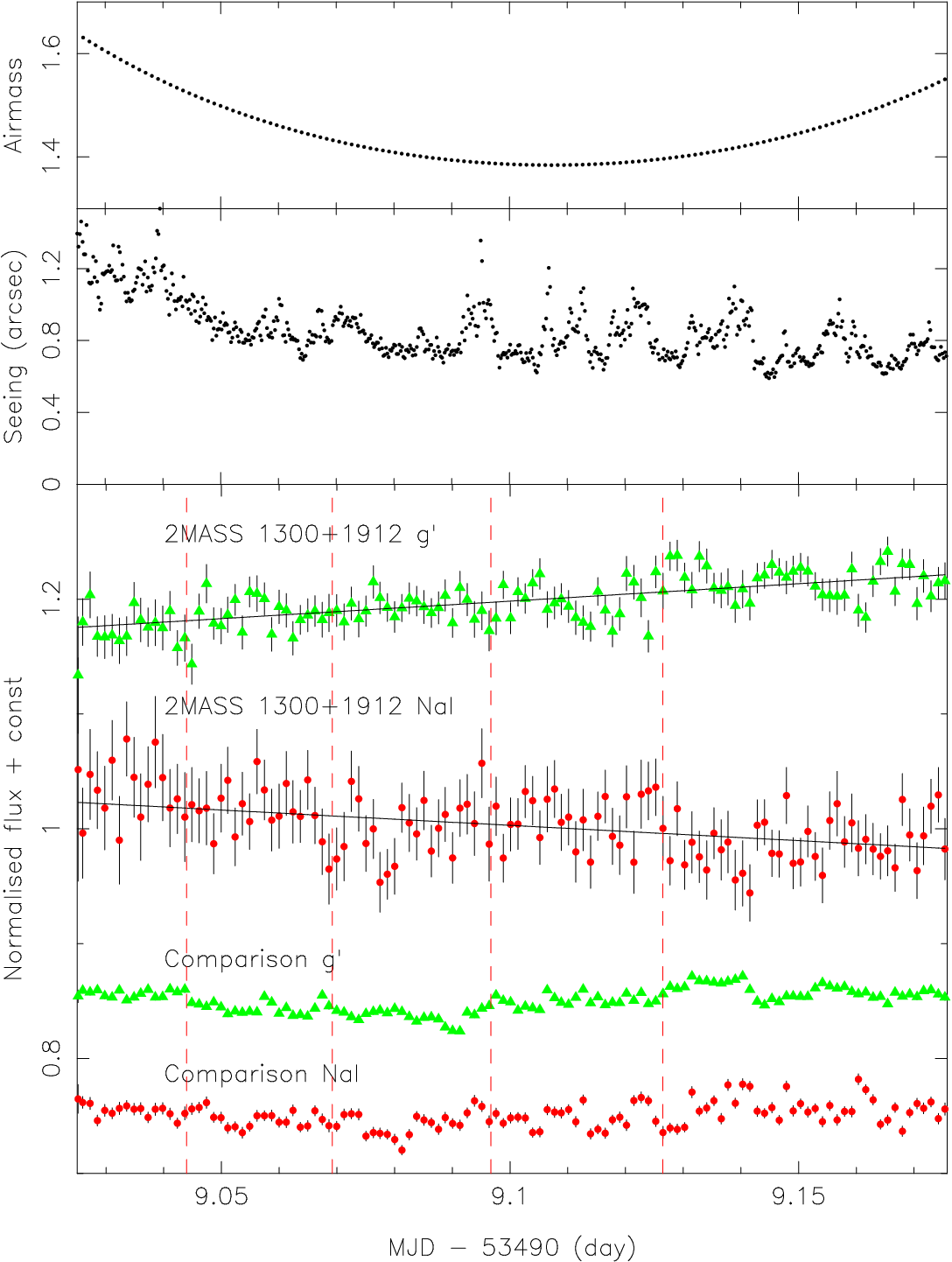}
\caption{Clockwise from top left: Lightcurves of 2MASS 2057-0252,
Kelu-1, 2MASS 1300+1912 and DENIS 1441-0945. The flux of the target
star, divided by a nearby comparison, is plotted against Modified
Julian Date. The lightcurves have been binned to a time resolution of
approximately one minute, normalised by division of the mean value,
and a vertical offset of 0.2 has been applied to the Sloan-$g'$
lightcurve.  The NaI lightcurve is plotted with red circles, the
Sloan-$g'$ lightcurve is plotted with green triangles. The lightcurves
of the comparison star, divided by a third, nearby comparison star are
plotted beneath the target star's lightcurves, with the same
symbols. These lightcurves have also been normalised and an offset
added for clarity. Also shown are 1$^{st}$ order polynomial fits to
the UCD data (except in the case of DENIS 1441-0945, where a 2$^{nd}$
order fit is shown). Dashed vertical lines in the 2MASS 1300+1912
panel show the times at which the star's position on the chip was
changed. For each object additional panels show the
seeing in arcsecs (middle panel) as measured from the FWHM of stellar
images, and the airmass of the UCD (top panel).}
\label{fig:lightcurves}
\end{center}
\end{figure*}

2MASS 2057-0252 is an L1.5 dwarf with an $I$-band magnitude of 16.5
\citep{koen03}. The object showed tentative evidence for $I$-band
variability on one night, which was absent on the second night it was
observed \citep{koen03}. Our observations of 2MASS 2057-0252 are
presented in figure~\ref{fig:lightcurves}. No evidence of variability
is present in the NaI lightcurve, as determined by a one-sided
F-test. Whilst the $g'$-band lightcurve shows a steady decline in flux
with time, the flux is strongly correlated with airmass, which
strongly suggests that this is due to a colour effect, arising because
the target is redder than the comparison star. This also naturally
explains the lack of variability in the NaI filter, as this filter is
much narrower, and redder, than the Sloan-$g'$ filter.  We therefore
conclude that this object shows no strong evidence for variability on
a 3 hour timescale.

\subsection{Kelu-1}
\label{subsec:kelu1}


\begin{figure}
\begin{center}
\includegraphics[scale=0.3,angle=-90]{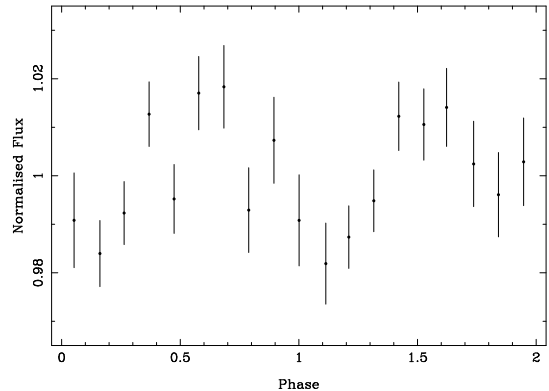}
\caption{The $g'$-band lightcurve of Kelu-1, phased on the known
  1.8-hr period, and binned into nineteen phase bins.  Data taken at high
  airmass and with a poor signal-to-noise ratio (after MJD 53504.22)
  was masked out prior to phasing.}
\label{fig:phased}
\end{center}
\end{figure}

Kelu-1 is an L2 dwarf \citep{ruiz97}, and a known binary star
\citep{liu05,gelino05}. Observations centred on the complex of molecular
bandheads at $\sim$8600\AA\, show strong periodic variability, with a
period of 1.8-hr \citep{clarke02b}. Time-resolved spectroscopy of the
object showed no evidence for variability in the dust-sensitive TiO
bandheads, however \citep{clarke03}. Our observations are presented in
figure~\ref{fig:lightcurves}. The NaI lightcurve shows no
evidence of variability: we used a one-sided F-test to show that a
1$^{st}$-order polynomial was a better match to the data than a
constant fit at a significance of just 50\%.

 The $g'$-band flux, however, shows a rising trend with time. This
rising trend is strongly correlated with airmass and is most likely a
colour effect.  Additional variability {\em is} visible, however, at
the start of the observations, where the airmass was low, and
signal-to-noise is better. A Lomb-Scargle periodogram of the $g'$-band
data taken before MJD 53504.22 (airmass 1.3) has a peak at the known
1.8-hr period of Kelu-1. Following the analysis of
\cite{schwarzenberg98}, the significance of this peak is better than
99.9 percent. The formal significance should be treated with caution;
non-white noise, deviations of the lightcurve from a pure sinusoid and
irregular sampling can all affect the periodogram, and the length of
our run is only just longer than the expected period. However, the
significance is high enough for us to conclude that the 1.8-hr
periodic variability at 8600\AA\, reported by \cite{clarke02b} is also
present in the $g'$-band light of Kelu-1. The phased lightcurve is
plotted in figure~\ref{fig:phased}.

\subsection{2MASS 1300+1912}
\label{subsec:2mass1300}

2MASS 1300+1912 is an L1 dwarf with an $I$-band magnitude of 15.9
\citep{gelino02}. Convincing evidence for non-periodic $I$-band
variability with an amplitude of $\sim$0.2 mags was reported by
\cite{gelino02}. Our observations are presented in
figure~\ref{fig:lightcurves}. The observations taken above airmass of
1.7 are not shown as the signal/noise ratio is too low to be
useful. It is clear from figure~\ref{fig:lightcurves} that the object
shows variability at a level of about 5 percent in both bands.  This
variability is significant, and clearly anti-correlated (the
correlation coefficient between the two lightcurves is $-0.09 \pm
0.01$). To determine the significance of the anti-correlation we
computed 2$^{nd}$ order polynomial fits to the lightcurves of the
target and comparison star. The difference between the $g'$-band and
NaI slopes is $-0.6\pm0.1$ and $0.00\pm0.02$ for the target star and
comparison star respectively. Futhermore, the individual slopes of the
comparison star's lightcurves are both consistent with zero. We
conclude that the anti-correlation is significant, and not introduced
by variability in the comparison star. Although the $g'$-band
lightcurves of the other objects are affected by colour-term effects,
this is not the case here. Whilst the airmass first falls and then
rises, the $g'$-band lightcurve of 2MASS 1200+1912 shows a steady
rising trend. We conclude that colour-term effects {\em do not} affect
the $g'$-band lightcurve of 2MASS 1300+1912. Furthermore, the
variability is not correlated with the seeing, sky brightness or
transparency, and we conclude that systematic errors resulting from
the removal and/or correction of these factors do not cause the
observed variability.

Such variability, if real, would be strong evidence that the
variability in this system is caused by dust clouds.  Unfortunately,
the position of the target on the CCD was not constant throughout the
run.  As discussed in section~\ref{sec:obs}, this could lead to
spurious variability in the lightcurve. A reasonable check of the
presence of such effects would be the structure in the differential
lightcurve of our chosen comparison and another star on the CCD. Only
one such star is present, approximately 530 pixels away from our
comparison star. The differential lightcurve of our comparison star
divided by this second star is plotted in
figure~\ref{fig:lightcurves}. Variability is indeed present, but only
at a level of approximately 1\%. Futhermore, there are no sudden jumps
in the lightcurves at the times when the star's position was
changed. We conclude that vignetting has not introduced features into
the lightcurve, although the variability in the lightcurve at the 1\%
level suggests that one of the comparison stars is itself variable,
albeit at a low level.

We therefore conclude that the anti-correlated variability seen in 2MASS
1300+1912 is real.  This conclusion is strengthened by the fact that
most systematic errors would produce spurious variability that is
{\em correlated} between the two wavebands. In particular, this
applies to both colour effects, and any errors introduced by the
uncorrected vignetting (the vignetting patterns are very similar in
both CCDs).

\subsection{DENIS 1441-0945}
\label{subsec:denis1441}

DENIS1441-0945 is an L1 dwarf with an $I$-band magnitude of 16.9
\citep{koen03}, and a known binary \citep{bouy03}.  It was observed in
the $I$-band by \cite{koen03} - no convincing evidence for variability
was reported. Although a rising trend was present in the lightcurve,
this may have been a colour-term effect.  Our observations of
DENIS1441-0945 are presented in figure~\ref{fig:lightcurves}. No
evidence of variability is present in the NaI lightcurve, as
determined by a one-sided F-test. Whilst the $g'$-band lightcurve
shows a gradual change in flux with time, the flux is strongly
correlated with airmass, which suggests that this is due to a
colour-term effect. We therefore conclude that this object shows no
evidence for variability.

\section{Discussion}
\label{sec:disc}

The detection of anti-correlated variability between the $g'$-band and
the NaI band in 2MASS 1300+1912 is incompatible with models of
starspot-induced variability. Given that anti-correlated variability
is expected from models of dust-cloud variability, and that rapid
evolution of dust clouds is to be expected in UCDs, it is likely that
dust clouds are responsible for the variability in this system. The
relative amplitude of variability between the two bands can constrain
the nature of the cloud evolution responsible for variability. The
models used in figure~\ref{fig:model} show that a dust cloud forming
on a predominantly clear amosphere would produce a much stronger
signal in the NaI band than the Sloan-$g'$ band. In contrast, a small
clearing in a predominantly dusty atmosphere produces variability of
about the same magnitude in each band. The amplitude of variability
can also constrain the size of the cloud or clearing; our observations
are consistent with the development of a clear patch covering
approximately 4 percent of a predominantly dusty atmosphere.

The result presented here demonstrates that simultaneous, multi-colour
photometry is effective in distinguishing between different
variability scenarios for UCDs, and opens up the possibility of
understanding the variability of UCDs in more detail. Clearly, the
method should be applied to a large sample of UCDs to determine if the
variability in {\em all} UCDs is due to dust clouds, and at what
effective temperatures dust-cloud variability becomes important.

\section{Conclusions}
\label{sec:conc}
We present simultaneous, multi-colour photometry of four ultracool
dwarfs with the triple-beam photometer {\sc ultracam}. Data were
obtained simultaneously in the Sloan-$g'$ band and a specially
designed narrow-band NaI filter. Of the four objects, only 2MASS
1300+1912 shows good evidence for variability in both bands. For this
object, the variability is anti-correlated at a high level of
significance, providing the first direct evidence that dust-cloud
weather is responsible for the variability in ultracool dwarfs.

\section*{\sc Acknowledgements}
TRM acknowledges the support of a PPARC Senior Research
Fellowship. ULTRACAM and SPL are supported by PPARC grants
PP/D002370/1 and PPA/G/S/2003/00058, respectively. TS acknowledges
support from the Spanish Ministry of Science and Technology under the
programme Ram\'{o}n Y Cajal.  The authors would like to thank Tim
Naylor for useful discussions.

\bibliographystyle{mn2e}
\bibliography{abbrev,refs,refs2}

\begin{thebibliography}{}

\bibitem[\protect\citeauthoryear{{Allard}, {Hauschildt}, {Alexander}, {Tamanai}
  \& {Schweitzer}}{{Allard} et~al.}{2001}]{allard01}
{Allard} F.,  {Hauschildt} P.~H.,  {Alexander} D.~R.,  {Tamanai} A.,
  {Schweitzer} A.,  2001, \apj, 556, 357

\bibitem[\protect\citeauthoryear{{Bailer-Jones}}{{Bailer-Jones}}{2002}]{bailer%
-jones02}
{Bailer-Jones} C.~A.~L.,  2002, \aap, 389, 963

\bibitem[\protect\citeauthoryear{{Bailer-Jones}}{{Bailer-Jones}}{2004}]{bailer%
-jones04}
{Bailer-Jones} C.~A.~L.,  2004, astro-ph/0409463

\bibitem[\protect\citeauthoryear{{Bailer-Jones} \& {Mundt}}{{Bailer-Jones} \&
  {Mundt}}{2001}]{bailer-jones01}
{Bailer-Jones} C.~A.~L.,  {Mundt} R.,  2001, \aap, 367, 218

\bibitem[\protect\citeauthoryear{{Bouy}}{{Bouy}}{2004}]{bouy04}
{Bouy} H.,  2004, \aap, 424, 619

\bibitem[\protect\citeauthoryear{{Bouy}, {Brandner}, {Mart{\'{\i}}n},
  {Delfosse}, {Allard} \& {Basri}}{{Bouy} et~al.}{2003}]{bouy03}
{Bouy} H.,  {Brandner} W.,  {Mart{\'{\i}}n} E.~L.,  {Delfosse} X.,  {Allard}
  F.,    {Basri} G.,  2003, \aj, 126, 1526

\bibitem[\protect\citeauthoryear{{Clarke}, {Oppenheimer} \& {Tinney}}{{Clarke}
  et~al.}{2002}]{clarke02a}
{Clarke} F.~J.,  {Oppenheimer} B.~R.,    {Tinney} C.~G.,  2002, \mnras, 335,
  1158

\bibitem[\protect\citeauthoryear{{Clarke}, {Tinney} \& {Covey}}{{Clarke}
  et~al.}{2002}]{clarke02b}
{Clarke} F.~J.,  {Tinney} C.~G.,    {Covey} K.~R.,  2002, \mnras, 332, 361

\bibitem[\protect\citeauthoryear{{Clarke}, {Tinney} \& {Hodgkin}}{{Clarke}
  et~al.}{2003}]{clarke03}
{Clarke} F.~J.,  {Tinney} C.~G.,    {Hodgkin} S.~T.,  2003, \mnras, 341, 239

\bibitem[\protect\citeauthoryear{{Dhillon} \& {Marsh}}{{Dhillon} \&
  {Marsh}}{2001}]{dhillon01}
{Dhillon} V.,  {Marsh} T.,  2001, New Astronomy Review, 45, 91

\bibitem[\protect\citeauthoryear{{Fleming}, {Giampapa} \& {Schmitt}}{{Fleming}
  et~al.}{2000}]{fleming02}
{Fleming} T.~A.,  {Giampapa} M.~S.,    {Schmitt} J.~H.~M.~M.,  2000, \apj, 533,
  372

\bibitem[\protect\citeauthoryear{{Gelino}, {Kulkarni} \& {Stephens}}{{Gelino}
  et~al.}{2005}]{gelino05}
{Gelino} C.~R.,  {Kulkarni} S.~R.,    {Stephens} D.~C.,  2005, ArXiv
  Astrophysics e-prints

\bibitem[\protect\citeauthoryear{{Gelino}, {Marley}, {Holtzman}, {Ackerman} \&
  {Lodders}}{{Gelino} et~al.}{2002}]{gelino02}
{Gelino} C.~R.,  {Marley} M.~S.,  {Holtzman} J.~A.,  {Ackerman} A.~S.,
  {Lodders} K.,  2002, \apj, 577, 433

\bibitem[\protect\citeauthoryear{{Herbst}, {Herbst}, {Grossman} \&
  {Weinstein}}{{Herbst} et~al.}{1994}]{herbst94}
{Herbst} W.,  {Herbst} D.~K.,  {Grossman} E.~J.,    {Weinstein} D.,  1994, \aj,
  108, 1906

\bibitem[\protect\citeauthoryear{{Koen}}{{Koen}}{2003}]{koen03}
{Koen} C.,  2003, \mnras, 346, 473

\bibitem[\protect\citeauthoryear{{Leggett}, {Allard} \& {Hauschildt}}{{Leggett}
  et~al.}{1998}]{leggett98}
{Leggett} S.~K.,  {Allard} F.,    {Hauschildt} P.~H.,  1998, \apj, 509, 836

\bibitem[\protect\citeauthoryear{{Liu} \& {Leggett}}{{Liu} \&
  {Leggett}}{2005}]{liu05}
{Liu} M.~C.,  {Leggett} S.~K.,  2005, \apj, 634, 616

\bibitem[\protect\citeauthoryear{{Mart{\'{\i}}n}, {Brandner}, {Bouvier},
  {Luhman}, {Stauffer}, {Basri}, {Zapatero Osorio} \& {Barrado y
  Navascu{\'e}s}}{{Mart{\'{\i}}n} et~al.}{2000}]{martin00}
{Mart{\'{\i}}n} E.~L.,  {Brandner} W.,  {Bouvier} J.,  {Luhman} K.~L.,
  {Stauffer} J.,  {Basri} G.,  {Zapatero Osorio} M.~R.,    {Barrado y
  Navascu{\'e}s} D.,  2000, \apj, 543, 299

\bibitem[\protect\citeauthoryear{{Mart{\'{\i}}n}, {Zapatero Osorio} \&
  {Lehto}}{{Mart{\'{\i}}n} et~al.}{2001}]{martin01}
{Mart{\'{\i}}n} E.~L.,  {Zapatero Osorio} M.~R.,    {Lehto} H.~J.,  2001, \apj,
  557, 822

\bibitem[\protect\citeauthoryear{{Mohanty}, {Basri}, {Shu}, {Allard} \&
  {Chabrier}}{{Mohanty} et~al.}{2002}]{mohanty02}
{Mohanty} S.,  {Basri} G.,  {Shu} F.,  {Allard} F.,    {Chabrier} G.,  2002,
  \apj, 571, 469

\bibitem[\protect\citeauthoryear{{Naylor}}{{Naylor}}{1998}]{naylor98}
{Naylor} T.,  1998, \mnras, 296, 339

\bibitem[\protect\citeauthoryear{{Ozawa}, {Grosso} \& {Montmerle}}{{Ozawa}
  et~al.}{2005}]{ozawa05}
{Ozawa} H.,  {Grosso} N.,    {Montmerle} T.,  2005, \aap, 429, 963

\bibitem[\protect\citeauthoryear{{Rockenfeller}, {Bailer-Jones} \&
  {Mundt}}{{Rockenfeller} et~al.}{2006}]{rockenfeller06a}
{Rockenfeller} B.,  {Bailer-Jones} C.~A.~L.,    {Mundt} R.,  2006, \aap, 448,
  1111

\bibitem[\protect\citeauthoryear{{Ruiz}, {Leggett} \& {Allard}}{{Ruiz}
  et~al.}{1997}]{ruiz97}
{Ruiz} M.~T.,  {Leggett} S.~K.,    {Allard} F.,  1997, \apjl, 491, L107+

\bibitem[\protect\citeauthoryear{{Schwarzenberg-Czerny}}{{Schwarzenberg-Czerny%
}}{1998}]{schwarzenberg98}
{Schwarzenberg-Czerny} A.,  1998, \mnras, 301, 831

\bibitem[\protect\citeauthoryear{{Tinney} \& {Tolley}}{{Tinney} \&
  {Tolley}}{1999}]{tinney99}
{Tinney} C.~G.,  {Tolley} A.~J.,  1999, \mnras, 304, 119

\end{thebibliography}

\end{document}